\documentclass[lettersize, journal]{IEEEtran}
\usepackage{amsmath,amsfonts}
\usepackage{algorithmic}
\usepackage{algorithm} 
\usepackage{array}
\usepackage[caption=false,font=normalsize,labelfont=sf,textfont=sf]{subfig}
\usepackage{textcomp}
\usepackage{stfloats}
\usepackage{url}
\usepackage{verbatim}
\usepackage{graphicx}
\usepackage{cite}
\usepackage{bm}                          
\usepackage{microtype}                   
\usepackage[colorlinks, linkcolor=blue, anchorcolor=blue, citecolor=blue]{hyperref} 
\usepackage{orcidlink}                   
\usepackage{amssymb}                     
\graphicspath{{fig/}}                    
\usepackage{booktabs}                    
\usepackage{multirow}                    
\usepackage{makecell}                    
\usepackage{threeparttable}				 

\begin{document}

\bstctlcite{IEEEexample:ControlAuthorNumber} 	

\title{Sequential Detection-Based Iterative Blind Separation for Single-Channel Co-Frequency Signals} 					  
 
\author
{
	Heng Wang $^{\orcidlink{0000-0003-0618-1239}}$,  
	Peng Sun $^{\orcidlink{0000-0001-6431-3119}}$,
	Kexian Gong $^{\orcidlink{0000-0003-4046-0426}}$, 
	Kunheng Zou $^{\orcidlink{0000-0002-3408-4543}}$,  
	and Hua Jiang $^{\orcidlink{0000-0002-4176-6833}}$
	
	\thanks{ (\emph{Corresponding author: Kexian Gong.}) }  
	\thanks{
		Heng Wang, 
		Peng Sun,  
		Kexian Gong,
		Kunheng Zou,
		and Hua Jiang
		are with the School of Electrical and Information Engineering, Zhengzhou University, Zhengzhou 450001, China 
		(e-mail: 
		w1270325699@gs.zzu.edu.cn;  
		iepengsun@zzu.edu.cn; 
		kxgong@zzu.edu.cn;
		zoukunheng@gs.zzu.edu.cn; 
		hjiang0619@sina.com).
	}
}    


\maketitle    
 
\begin{abstract}
	Existing single-channel co-frequency signal blind separation (SCSBS) algorithms struggle to balance separation accuracy, computational complexity, and robustness, while current channel state information (CSI) estimation methods lack precision. 
	To address these limitations, we propose a sequential detection (SD)-based iterative separation (SDIS) algorithm. 
	SDIS incorporates a delayed unscented Kalman filter (DUKF) into an iterative decision feedback framework, jointly enhancing signal separation and CSI estimation.
	Simulation results show that SDIS outperforms benchmarks in separation accuracy, CSI estimation accuracy, computational efficiency, and robustness. 
	Notably, when the mean bit error rate (MBER) drops below $10^{-4}$, SDIS can tolerate at least $0.8$ dB more noise than the benchmarks.
\end{abstract}		 
 
\begin{IEEEkeywords}
	Single-channel co-frequency signal, sequential detection, iterative separation, delayed unscented Kalman filter.
\end{IEEEkeywords}

\section{Introduction}
\label{Sec_js}    
As spectrum resources become increasingly scarce, co-frequency interference occurs frequently~\cite{r_nhw_2025}, and time-frequency resource reuse techniques are being widely adopted~\cite{r_djz_2025}. 
Consequently, multiple independent signals (termed sub-signals) increasingly share the same frequency band simultaneously, rendering co-frequency signals ever more prevalent~\cite{r_wh_2025}. 
Single-channel co-frequency signal (SCS) blind separation (SCSBS) has attracted considerable attention, as it blindly recovers all sub-signals from an SCS.

Per-survivor processing (PSP)-based algorithms dominate SCSBS~\cite{r_tsl_2008}, \cite{r_lxb_2017}, \cite{r_yy_2020} due to their superior performance over successive interference cancellation (SIC)-type~\cite{r_gwb_2022} and particle filter (PF)-type~\cite{r_gym_2019} algorithms.
The PSP combined with least mean squares (LMS) channel estimation is proposed in~\cite{r_tsl_2008} and termed PSPL.
The work in~\cite{r_lxb_2017} introduces an improved PSP algorithm with carrier frequency offset (CFO) pre-compensation, termed PCP.
To improve separation accuracy, a PSP variant, namely the two-stage Viterbi-like detection (TVD) algorithm, is proposed in~\cite{r_yy_2020}.
However, PSPL, PCP, and TVD suffer from high complexity when processing higher-order modulated signals.
To address this, the work in~\cite{r_my_2} presents a low-complexity sequential detection (SD)-based separation (SDS) method.

Since PSP-based methods and SDS are sensitive to channel state information (CSI) pre-estimation errors (CSIPEs) (i.e., errors from parameter pre-estimation), we propose a delayed unscented Kalman filter (DUKF) to mitigate these errors, thereby enabling blind estimation of key CSI parameters, namely channel gain (CG), CFO, carrier phase offset (CPO), and symbol timing offset (STO). 
For CSI estimation, CFO and STO estimators are presented in~\cite{r_fa_2020} and~\cite{r_fa_2021}, respectively, while joint estimation of CG and CPO is achieved in~\cite{r_lmq_2024}.
Nevertheless, the estimation accuracy of these methods is substantially inferior to that of the exhaustive-search-based four-parameter joint estimation (FPJE) algorithm in~\cite{r_gym_2018}. 
However, the performance of FPJE itself remains limited by its inherent search step size. 

In summary, existing SCSBS techniques struggle to balance accuracy, efficiency, and robustness, and current CSI estimation methods are insufficiently accurate. 
To address these issues, we propose an SD-based iterative separation (SDIS) algorithm. 
Our contributions are as follows:
\begin{itemize}
	\item[1)] SDIS uses SD for low-complexity symbol detection and integrates a DUKF for CSI estimation, thereby enhancing robustness to CSIPEs.
	
	\item[2)] SDIS employs iterative decision feedback to alternately update symbol decisions and CSI estimates, thereby refining both.
	
	\item[3)] Numerical results show that SDIS outperforms benchmarks in separation accuracy, CSI estimation accuracy, computational efficiency, and robustness, with gains increasing with modulation order.
\end{itemize}  

\emph{Notation}: For symbols $a$ and $b$, $(a, b)$ denotes an ordered pair.   
For a fixed integer $w$ and integers $c \leq d$, define $[x_{w + v}]_{v = c}^{d} \triangleq [x_{w + c}, x_{w + c + 1}, \dots, x_{w + d}]^\top$.

\section{System Model}
\label{Sec_xtmx}  
For terminal $i \in \{0, 1\}$, the bit sequence is mapped onto the $M^i$-ary constellation $\mathcal{S}^i$ to yield the symbol sequence $\bm{s}^i$, which is then pulse-shaped to form the sub-signal $\bm{r}^i$. 
With identical symbol rates and closely spaced carrier frequencies, $\bm{r}^0$ and $\bm{r}^1$ overlap in the time-frequency domain and are superimposed to produce the co-frequency signal. 
After transmission and matched filtering over an additive white Gaussian noise (AWGN) channel, the observed SCS $\bm{y}$ is obtained. 
The $n$-th sample of $\bm{y}$ is $y_n = \gamma_n + v_n$, where the noise-free signal $\gamma_n$ is
\begin{subequations}
	\label{Eq_ygfs}
	\begin{align}   
		\gamma_n 
		& = \sum\limits_{i = 0}^{1}{h^i e^{j \left( 2 \pi \theta_n^i + \varphi^i \right)}} \mathcal{A}_n^i , \\		 
		\mathcal{A}_n^i 
		& = \sum\limits_{m = -L_0^i}^{L_1^i} s_{q + m}^i g^i \left( ( n - q p ) T_s - mT - \varepsilon^iT \right) ,
	\end{align}
\end{subequations}  
$v_n \sim \mathcal{CN}(0, N_0)$ denotes complex AWGN with a one-sided power spectral density of $N_0$, $n \in \{ 0, 1, \dots, N - 1 \}$, and $N$ is the total number of samples.
Moreover, $T_s$ is the sampling period, $T = p T_s$ is the symbol period with oversampling factor $p$, $q = \lfloor n / p \rfloor$ is the symbol index, and $s_q^i$ is the $q$-th symbol of $\bm{s}^i$.
The parameters $h^i$, $f^i$, $\varphi^i$, and $\varepsilon^i$ denote the CG, symbol-rate-normalized CFO, CPO, and STO of $\bm{r}^i$, respectively, and $\theta_n^i = f^i n / p$ represents the normalized accumulated CFO phase.
The pulse $g^i(\cdot)$ is the convolution of the shaping and matched filter responses for $\bm{r}^i$, with $L^i = L_0^i + L_1^i + 1$, where $L_0^i$ and $L_1^i$ count the causal and non-causal symbols, respectively. 

Without loss of generality, we set $p = 1$, $M = M^0 = M^1$, $\mathcal{S} = \mathcal{S}^0 = \mathcal{S}^1$, $L_0 = L_0^0 = L_0^1$, $L_1 = L_1^0 = L_1^1$, and $L = L^0 = L^1$.
The symbols are equiprobable phase-shift keying (PSK) symbols, shaped by a raised-cosine pulse $g(\cdot) = g^0(\cdot) = g^1(\cdot)$ with roll-off factor $\alpha$.
The CSI parameters are constant over the observation interval, and the symbols are mutually independent.

\section{The SDIS Algorithm}

\subsection{Key Innovations of SDIS}

\begin{figure}[t]   
	\centering
	\includegraphics[width = 88 mm]{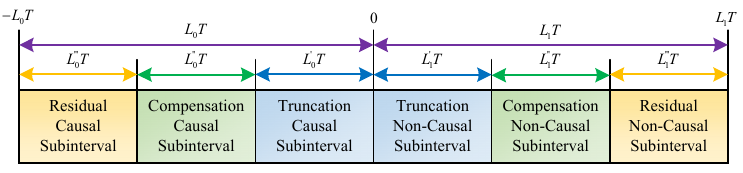}   
	\caption{Segmentation of the time-domain support region for $g(\cdot)$.}
	\label{Fig_RangeL}   
\end{figure} 

In the SD framework \cite{r_my_2}, the candidate path (CP) with the largest path metric (PM) yields the most reliable decision symbol sequence.  
However, calculating PM requires the estimation of $\gamma_n$, which involves $L$ symbols ($L \to \infty$), making it necessary to truncate $L$ in practice.
As shown in Fig.~\ref{Fig_RangeL}, we exploit the sharp main-lobe property of the raised-cosine pulse~\cite{r_by_2025} to partition $g(\cdot)$ of length $LT$ into three types of subintervals: truncation ($L_0'T, L_1'T$), compensation ($L_0''T, L_1''T$), and residual ($L_0'''T, L_1'''T$).
By neglecting the minor effect of the residual part, we obtain the estimator $\hat{\gamma}_n$ of $\gamma_n$ as 
\begin{subequations}
	\label{Eq_hGnhGnc}
	\begin{align}   
		\label{Eq_hGnhGnc_1}
		\hat{\gamma}_n 
		& = \sum\limits_{c = 0}^{2} \hat{\gamma}_{n, c}, \quad c \in \left\{ 0, 1, 2 \right\} , \\
		\label{Eq_hGnhGnc_2}
		\hat{\gamma}_{n, c} 
		& = \sum \limits_{i = 0}^1 {\left\{ \hat{h}_n^i e^{j \left( 2 \pi \hat{\theta}_n^i + \hat{\varphi}_n^i \right)} \sum \limits_{l \in {\bm{l}_c}} \left[ \hat{\phi}_{n, l}^i \cdot \hat{s}_{n + l}^{i} \right] \right\} } , \\
		\label{Eq_hGnhGnc_3}
		\hat{\phi}_{n, l}^i 
		& = \frac{\sin \left[ \left(l + \hat{\varepsilon}_n^i \right)\pi \right]}{\left( l + \hat{\varepsilon}_n^i \right)\pi} \cdot \frac{\cos \left[ \pi \alpha \left( l + \hat{\varepsilon}_n^i \right) \right]}{1 - 4\alpha ^2{\left( l + \hat{\varepsilon}_n^i \right)}^2} ,
	\end{align}
\end{subequations}  
where $\bm{l}_0 = \{ -L_0'' - L_0', -L_0'' - L_0' + 1, \dots, -L_0' - 1 \}$, $\bm{l}_1 = \{ -L_0', -L_0' + 1, \dots, L_1' \}$, and $\bm{l}_2 = \{ L_1' + 1, L_1' + 2, \dots, L_1' + L_1'' \}$.
Here, $\hat{s}_n^i$, $\hat{h}_n^i$, $\hat{\varphi}_n^i$, $\hat{\varepsilon}_n^i$, and $\hat{\theta}_n^i$ denote the estimates of $s_n^i$, $h^i$, $\varphi^i$, $\varepsilon^i$, and $\theta_n^i$, respectively.     

In SDS~\cite{r_my_2}, since $\hat{s}_n^i$ are initially unknown, the PM $\psi$ of a CP is approximated as $\psi = \sum_{n=0}^{N-1}\beta_n$, where $\beta_n$, the branch metric (BM) of the $n$-th branch, is expressed as
\begin{subequations}
	\label{Eq_fjggp}
	\begin{align}  
		\label{Eq_fjggp_2}
		\beta_n 
		& \approx -\frac{| y_n - \hat{\gamma}_{n, 1} |^2}{N_0} + \lambda \left( 1 + \frac{\overline{\gamma}_{n, 0} + \overline{\gamma}_{n, 2}}{N_0} \right) , \\		 
		\overline{\gamma}_{n, c} 
		& = \sum\limits_{i = 0}^1 \left\{ \left| \hat{h}_n^i \right|^2 \sum \limits_{l \in \bm{l}_c} \left| \hat{\phi}_{n, l}^i \right|^2 \right\} .
	\end{align}
\end{subequations}   
Here, $|\cdot|$ denotes the complex modulus and $\lambda$ is a compensation factor. 
However, the approximate compensation terms $\overline{\gamma}_{n,0}$ and $\overline{\gamma}_{n,2}$ cannot accurately compensate for the effects of $\hat{\gamma}_{n,0}$ and $\hat{\gamma}_{n,2}$ on $\beta_n$, leading to degraded estimation accuracy. 
To address this, we propose an iterative decision feedback mechanism for real-time estimation of $\hat{\gamma}_{n,0}$ and $\hat{\gamma}_{n,2}$, thereby improving the accuracy of $\beta_n$.
Specifically, let $u \geq 0$ denote the iteration index of SDIS.
For $u = 0$, $\beta_n$ is given by \eqref{Eq_fjggp}.
When $u > 0$, by leveraging the exact form of $\beta_n$~\cite{r_my_2} and $\hat{s}_n^i$ from iteration $u - 1$, we reformulate $\beta_n$ as
\begin{equation}
	\beta_n 
	\approx -\frac{ \left| y_n - \hat{\gamma}_{n, 0}' - \hat{\gamma}_{n, 1} - \hat{\gamma}_{n, 2}' \right|^2}{N_0} + \lambda ,
	\label{Eq_mJn}
\end{equation} 
where the estimates $\hat{\gamma}_{n, 0}'$ and $\hat{\gamma}_{n, 2}'$ are obtained by substituting $[ \hat{s}_{n + l}^i ]_{l = -L_0'' - L_0'}^{-L_0' - 1}$ and $[ \hat{s}_{n + l}^i ]_{l = L_1' + 1}^{L_1' + L_1''}$ into \eqref{Eq_hGnhGnc_2}, respectively.
For notational simplicity, define $L' = L_0' + L_1' + 1$ and $L'' = L' + L_0'' + L_1''$.
 
On the other hand, CSIPEs render $\hat{h}_n^i, \hat{\varphi}_n^i, \hat{\varepsilon}_n^i$, and $\hat{\theta}_n^i$ unreliable, thus impairing the accuracy of $\beta_n$. 
Therefore, we discard the LMS method used in the benchmarks~\cite{r_tsl_2008}, \cite{r_lxb_2017}, \cite{r_yy_2020} due to its poor performance in nonlinear systems, and instead propose a DUKF for accurate tracking of these parameters. 
For the unscented Kalman filter fundamentals, see~\cite{r_lhy_2025}; we describe only the key improvements. 
Let $\bm{x}_n = [ \hat{h}_n^0, \hat{f}_n^0, \hat{\varphi}_n^0, \hat{\varepsilon}_n^0, \hat{\theta}_n^0, \hat{h}_n^1, \hat{f}_n^1, \hat{\varphi}_n^1, \hat{\varepsilon}_n^1, \hat{\theta}_n^1 ]^\top$ denote the $n$-th state vector, where $\hat{f}_n^i$ denotes the estimate of $f^i$. 
The state transition is then expressed as   
\begin{equation}   
	\bm{x}_{n + 1}  
	= \left[ \hat{h}_n^0, \hat{f}_n^0, \hat{\varphi}_n^0, \hat{\varepsilon}_n^0, \hat{\theta}_n^0 + \hat{f}_n^0, \hat{h}_n^1, \hat{f}_n^1, \hat{\varphi}_n^1, \hat{\varepsilon}_n^1, \hat{\theta}_n^1 + \hat{f}_n^1 \right]^\top ,
	\label{Eq_Axn}   
\end{equation}  
and the observation equation is given in \eqref{Eq_hGnhGnc}.
Updating $\bm{x}_{n + 1}$ requires $[\hat{\bm{s}}_{n + l}]_{l = -L_0' - L_0''}^{L_1' + L_1''}$, with $\hat{\bm{s}}_n \triangleq ( \hat{s}_n^0, \hat{s}_n^1 )$.
However, the newest $[\hat{\bm{s}}_{n + l}]_{l = -L_0'' - L_0'}^{L_1'}$ is available only when computing $\beta_n$, while the remaining part $[\hat{\bm{s}}_{n + l}]_{l = L_1' + 1}^{L_1' + L_1''}$ is accessible only after computing $\beta_{n + L_1''}$.
To eliminate a timing mismatch between the SD search and state updates, we update $\bm{x}_{n + L_1''}$ after reaching the $(n + L_1'')$-th branch, using $[\hat{\bm{s}}_{n + l}]_{l = -L_0' - L_0''}^{L_1' + L_1''}$ and $y_n$.
Since $L_1''$ is small~\cite{r_my_2} and $\theta_{n + L_1''}^i$ is estimated independently, this short delay barely affects convergence.
Finally, to mitigate stochastic fluctuations and guarantee asymptotic consistency, the final estimates of $h^i$, $f^i$, $\varphi^i$, and $\varepsilon^i$ are obtained via ensemble averaging over the converged samples, as follows:
\begin{equation}
	\label{Eq_hfvv}
	\begin{aligned} 
		\bar{h}^i
		& = \frac{2}{N} \sum_{n = N / 2}^{N - 1} \hat{h}_n^i , \qquad
		\bar{f}^i
		= \frac{2}{N} \sum_{n = N / 2}^{N - 1} \hat{f}_n^i , \\ 
		\bar{\varphi}^i
		& = \frac{2}{N} \sum_{n = N / 2}^{N - 1} \hat{\varphi}_n^i , \qquad 	
		\bar{\varepsilon}^i
		= \frac{2}{N} \sum_{n = N / 2}^{N - 1} \hat{\varepsilon}_n^i .
	\end{aligned}	 
\end{equation}

\subsection{Workflow of SDIS}
\label{Sec_SDIS_sflc}  
 
\begin{figure}[t]   
	\centering
	\includegraphics[width = 88 mm]{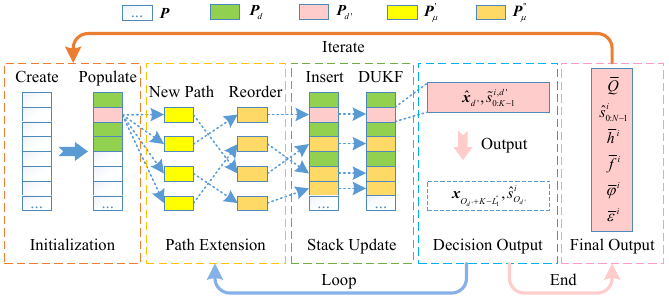}   																
	\caption{Workflow of the SDIS algorithm.}
	\label{Fig_SDIS}   
\end{figure}

SDIS integrates DUKF-based CSI estimation with iterative decision feedback to progressively refine both the CSI estimates and symbol decisions.
The workflow of SDIS is shown in Fig.~\ref{Fig_SDIS} and detailed below.

\emph{Step 1) Iteration Initialization.}  
SDIS maintains a stack $\bm{P}$ of $D \geq M^2$ rows, indexed by $d \in \{0, 1, \dots, D - 1\}$, each storing a windowed CP (WCP) $\bm{P}_d$ with at most $K$ symbol pairs \cite{r_my_2}.
Let $(\tilde{s}_{0 : K_d - 1}^{0, d}, \tilde{s}_{0 : K_d - 1}^{1, d})$ be the sequence of symbol pairs in $\bm{P}_d$, where $K_d \in \{0, 1, \cdots, K\}$ is the number of stored pairs.
Define $(\eta_\mu^0, \eta_\mu^1) \in \mathcal{S} \times \mathcal{S}$ as a candidate symbol pair with $\mu \in \{0, 1, \dots, M^2 - 1\}$.
Let $U$ be the maximum number of iterations, set the loop counter $Q = 0$, and initialize $u = 0$.

\emph{Step 2) Loop Initialization.}  
At the beginning of each iteration, clear $\bm{P}$.
For $d < M^2$, set $(\tilde{s}_0^{0, d}, \tilde{s}_0^{1, d}) = (\eta_d^0, \eta_d^1)$.
For every $d < D$, assign
\begin{equation}
	\left( K_d, \psi_d, O_d \right) =
	\begin{cases}
		\left( 1, 0, 0 \right),  		& \text{if } d < M^2 , \\
		\left( 0, -\infty, 0 \right), 	& \text{otherwise} ,
	\end{cases}
	\label{Eq_KFM} 
\end{equation} 
where $\psi_d$ is the PM of $\bm{P}_d$ and $O_d$ is the number of decisions already output from $\bm{P}_d$.
Let $d' = 0$ and $d'' = M^2$ denote the indices of the WCPs with the maximum and minimum PMs, respectively.
Initialize the state vector associated with $\bm{P}_d$ to $\hat{\bm{x}}_d = [ \bar{h}^0, \bar{f}^0, \bar{\varphi}^0, \bar{\varepsilon}^0, 0, \bar{h}^1, \bar{f}^1, \bar{\varphi}^1, \bar{\varepsilon}^1, 0 ]$, where $\bar{h}^i$, $\bar{f}^i$, $\bar{\varphi}^i$, and $\bar{\varepsilon}^i$ are obtained via any pre-estimation algorithm at $u = 0$ \cite{r_gym_2018}.

\emph{Step 3) Path Extension.} 
Increment $Q$ and branch $\bm{P}_{d'}$ into $M^2$ new WCPs $\bm{P}_\mu'$ using $(\eta_\mu^0, \eta_\mu^1)$, updating the symbol pair sequence of $\bm{P}_\mu'$ to $(\eta_{0 : K_{d'}}^{0, \mu}, \eta_{0 : K_{d'}}^{1, \mu}) = ([\tilde{s}_{0 : K_{d'} - 1}^{0, d'}, \eta_\mu^0], [\tilde{s}_{0 : K_{d'} - 1}^{1, d'}, \eta_\mu^1])$.  
The PM of $\bm{P}_\mu'$ is given by 
\begin{equation}
	\psi_\mu' = \psi_{d'} + \beta_\mu',
	\label{Eq_mFm}
\end{equation}
where the BM $\beta_\mu'$ of this extension is computed as follows: if $u = 0$, substitute $\eta_{K_{d'} - L' + 1 : K_{d'}}^{i, d'}$ and $\hat{\bm{x}}_{d'}$ into \eqref{Eq_fjggp_2}; if $u > 0$, substitute $\hat{s}_{O_{d'} + K_{d'} + 1 : O_{d'} + K_{d'} + L_1''}^i$, $\eta_{K_{d'} - L' + 1 : K_{d'}}^{i, d'}$, and $\hat{\bm{x}}_{d'}$ into \eqref{Eq_mJn}.
Finally, sort $\psi_{0 : M^2 - 1}'$ in ascending order to obtain $\psi_{0 : M^2 - 1}''$, and permute $\bm{P}_{0 : M^2 - 1}'$ accordingly into $\bm{P}_{0 : M^2 - 1}''$.

\emph{Step 4) Stack Update.}
Set $\bm{P}_{d'} = \bm{P}_{M^2 - 1}''$ and $K_{d'} = K_{d'} + 1$.
Feed $(\tilde{s}_{K_{d'} - L'' : K_{d'} - 1}^{0, d'}, \tilde{s}_{K_{d'} - L'' : K_{d'} - 1}^{1, d'})$ and $y_{O_{d'} + K_{d'} - L_1' - L_1'' - 2}$ into DUKF to update $\hat{\bm{x}}_{d'}$, and search the stack to refresh $d'$.
Then, for each $\mu < M^2 - 1$ with $\psi_{d''} < \psi_\mu''$, set $\bm{P}_{d''} = \bm{P}_\mu''$ and $K_{d''} = K_{d''} + 1$; feed $(\tilde{s}_{K_{d''} - L'' : K_{d''} - 1}^{0, d''}, \tilde{s}_{K_{d''} - L'' : K_{d''} - 1}^{1, d''})$ and $y_{O_{d''} + K_{d''} - L_1' - L_1'' - 2}$ into DUKF to update $\hat{\bm{x}}_{d''}$, and refresh $d''$ via stack search.        

\emph{Step 5) Cyclic Output.}
If $K_{d'} < K$, return to \emph{Step 3)}.  
Otherwise, output decisions and slide the window via
\begin{equation}
	\label{Eq_xs} 
	\bm{x}_{O_{d'} + K - L_1''} = \hat{\bm{x}}_{d'} , \qquad \hat{s}_{O_{d'}}^i = \tilde{s}_{0}^{i, d'} ,
\end{equation}
\begin{equation}
	\label{Eq_sKO} 
	\tilde{s}_{0 : K - 2}^{i, d'} = \tilde{s}_{1 : K - 1}^{i, d'} , \qquad
	K_{d'} = K - 1 , \qquad
	O_{d'} = O_{d'} + 1 ,
\end{equation} 
and proceed to \emph{Step 6)} if $O_{d'} + K_{d'} = N + L_1'$; otherwise, return to \emph{Step 3)}.

\emph{Step 6) Iterative Output.}
Update $\bar{h}^i$, $\bar{f}^i$, $\bar{\varphi}^i$, and $\bar{\varepsilon}^i$ via \eqref{Eq_hfvv} with $\bm{x}_{0 : N - 1}$, and increment $u$.
If $u = U$, compute the average loop number $\overline{Q} = Q / (UN)$ and terminate SDIS; otherwise, return to \emph{Step 2)}.

Algorithm~\ref{Alg_SDIS} summarizes the implementation of SDIS, whose numerical performance is evaluated next. 

\begin{algorithm}[t] 
	\caption{The SDIS Algorithm}
	\label{Alg_SDIS}
	\begin{algorithmic}[1]
		\REQUIRE $\bar{h}^i$, $\bar{f}^i$, $\bar{\varphi}^i$, and $\bar{\varepsilon}^i$.		
		\STATE Initialize stack $\bm{P}$, set $U$, $Q = 0$, and $u = 0$.
		\WHILE[\emph{Iteration}]{$u < U$}
		\STATE Clear $\bm{P}$. For $d < M^2$, set $( \tilde{s}_0^{0, d}, \tilde{s}_0^{1, d} ) = ( \eta_d^0, \eta_d^1 )$. For $d < D$, apply \eqref{Eq_KFM}.
		\STATE Set $d' = 0$, $d'' = M^2$, and initialize $\hat{\bm{x}}_d$.
		\WHILE[\emph{Loop}]{$O_{d'} + K_{d'} < N + L_1'$}
		\STATE $Q \leftarrow Q + 1$. For each $\mu < M^2$, extend $\bm{P}_{d'}$ to obtain $\bm{P}_\mu'$ and compute $\psi_\mu'$ using \eqref{Eq_fjggp_2}, \eqref{Eq_mJn}, and \eqref{Eq_mFm}.
		\STATE Sort $\psi_{0 : M^2 - 1}'$ ascending to obtain $\psi_{0 : M^2 - 1}''$, and permute $\bm{P}_{0 : M^2 - 1}'$ accordingly to yield $\bm{P}_{0 : M^2 - 1}''$.
		\STATE Set $\bm{P}_{d'} \leftarrow \bm{P}_{M^2 - 1}''$, $K_{d'} \leftarrow K_{d'} + 1$, update $\hat{\bm{x}}_{d'}$ via DUKF, and refresh $d'$.
		\STATE For each $\mu < M^2 - 1$, if $\psi_{d''} < \psi_\mu''$, set $\bm{P}_{d''} \leftarrow \bm{P}_\mu''$, $K_{d''} \leftarrow K_{d''} + 1$, update $\hat{\bm{x}}_{d''}$ via DUKF, and refresh $d''$.
		\STATE If $K_{d'} = K$, apply \eqref{Eq_xs} and \eqref{Eq_sKO} to output and slide the window.
		\ENDWHILE
		\STATE Apply \eqref{Eq_hfvv} to $\bm{x}_{0 : N - 1}$ to update $\bar{h}^i$, $\bar{f}^i$, $\bar{\varphi}^i$, and $\bar{\varepsilon}^i$. Then, set $u \leftarrow u + 1$.
		\ENDWHILE 
		\RETURN $\overline{Q} = Q / (UN)$, $\hat{s}_{0 : N - 1}^i$, $\bar{h}^i$, $\bar{f}^i$, $\bar{\varphi}^i$, and $\bar{\varepsilon}^i$.
	\end{algorithmic}
\end{algorithm}

\section{Numerical Results and Analysis}
\label{Sec_szjgfx} 
The evaluated algorithms include PSPL \cite{r_tsl_2008}, PCP \cite{r_lxb_2017}, TVD \cite{r_yy_2020}, SDS \cite{r_my_2}, FPJE \cite{r_gym_2018}, and SDIS.

\subsection{Experimental Setup}
\label{Sec_szjgfx_syhj}     
Let $E_s$ be the average symbol power of the SCS. 
The bit error rate (BER) and the normalized mean square error (NMSE) of CSI estimates, averaged over $\bm{r}^0$ and $\bm{r}^1$, are denoted as mean BER (MBER) and mean NMSE (MNMSE), respectively. 
The CSI pre-estimation errors are defined as $\Delta(h^i) = (\bar{h}^i - h^i)/h^i$, $\Delta(f^i) = \bar{f}^i - f^i$, $\Delta(\varphi^i) = (\bar{\varphi}^i - \varphi^i) / (2\pi)$, and $\Delta(\varepsilon^i) = \bar{\varepsilon}^i - \varepsilon^i$.
 
Simulations are conducted in MATLAB R2020a on Windows 10 (64-bit) with an Intel Core i5-12400F processor and 16 GB DDR4 RAM.  
Results are averaged over 100 Monte Carlo trials.
Each terminal transmits $24,000$ bits per trial and employs a quadrature PSK (QPSK) modulator. 
Here, we set $p = 1$, $\alpha = 0.35$, $E_s / N_0 = 19$ dB, CG ratio $h_r = h^1 / h^0 = 0.9$, $\varepsilon^0 = 0.05$, $\varepsilon^1 = 0.35$, and assume $f^i$ and $\varphi^i$ are uniformly distributed over $[-0.002, 0.002]$ and $[-\pi, \pi)$, respectively. 	 	
The CSI pre-estimation errors are set as $\Delta (h^0) = \Delta (\varphi^0) = \Delta (\varepsilon^0) = 0.01$, $\Delta (h^1) = \Delta (\varphi^1) = \Delta (\varepsilon^1) = -0.01$, and $\Delta (f^0) = -\Delta (f^1) = -10^{-5}$. 
In DUKF, the covariances for $\hat{h}_0^i$, $\hat{f}_0^i$, $\hat{\varphi}_0^i$, $\hat{\varepsilon}_0^i$, and $\hat{\theta}_0^i$ are set to $4\times10^{-4}(\bar{h}^i)^2$, $4\times10^{-10}$, $4\times10^{-4}$, $4\times10^{-4}$, and $4\times10^{-8}$, respectively; the scaling parameter is $0.1$, and other settings follow \cite{r_lhy_2025}.
Referring to~\cite{r_my_2}, for SDS (CSI perfectly known) and SDIS, we set $U = 2$, $L_0'' = 0$, $L_0' = 3$, $L_1' = 1$, $L_1'' = 2$, $\lambda = 1.5$, $K = 12$, and $D = 16M^2$. 
For other benchmarks, we set $L_0' = L_1' = 1$, $L_0'' = L_1'' = 2$, the LMS step size to $0.005$, and the traceback depth to $12$ symbols~\cite{r_yy_2020}. 
In particular, PSPL with $L_0' = 2$ is denoted PSPL4, FPJE adopts $p = 4$, and PCP is assumed to achieve CFO estimation accuracy of $10^{-5}$.

\subsection{Computational Complexity Analysis}
\label{Sec_szjgfx_fzdfx}   

\begin{table}[t] 
	\centering 
	\caption{Per-Sample Computational Complexity of the Methods}
	\label{Tab_fzd} 
	\begin{tabular}{l l l} 
		\toprule		 				  
		Method			  
		& Sequence Detection	 												  
		& CSI Estimation \\ 
		
		\midrule	 	
		PSPL \cite{r_tsl_2008}
		& $\mathcal{O} ( 2L' M^{2L'} )$ 						 	 		
		& $\mathcal{O} ( 2L' M^{2L'-2} )$ \\  
		
		PCP \cite{r_lxb_2017}
		& $\mathcal{O} ( 2L' M^{2L'} )$
		& $\mathcal{O} ( 2L' M^{2L' - 2} )$ \\  
		
		TVD \cite{r_yy_2020}
		& $\mathcal{O} ( 2(L' + L'') M^{2L'} )$ 					 		
		& $\mathcal{O} ( 2(L' + L'') M^{2L' - 2} )$ \\  
		
		SDIS
		& $\mathcal{O} ( 2(L' + (U - 1)L'')\overline{Q}M^2 )$
		& $\mathcal{O} ( L_x^3 U\overline{Q}M^2/2 )$  \\
		\bottomrule	
	\end{tabular} 
\end{table}
   
Table~\ref{Tab_fzd} divides the per-sample complexity into sequence detection and CSI estimation, where $L_x$ is the state dimension of DUKF.  
The complexity of PSPL, PCP, and TVD (per stage) is dominated by $M^{2L'}$ BM computations and $2M^{2L'-2}$ LMS updates \cite{r_tsl_2008}, \cite{r_lxb_2017}, \cite{r_yy_2020}.  
Each iteration of SDIS performs $\overline{Q}M^2$ BM computations (involving $2L'$ symbols for $u=0$ and $2L''$ for $u>0$) and approximately $\overline{Q}M^2/2$ DUKF operations.   
In summary, SDIS scales as $M^2$ versus $M^{2L'}$ for the benchmarks.
The running times of the algorithms in Table~\ref{Tab_sfhsyfl} confirm that the advantage of SDIS increases with $M$ and $L'$.

\subsection{Impact of Noise and Modulation on Separation} 
\label{Sec_szjgfx_zsyfl} 
\begin{figure}[t]
	\centering   
	\subfloat[][\label{Fig_sy_zsyfl_a}]{\includegraphics[width = 43 mm]{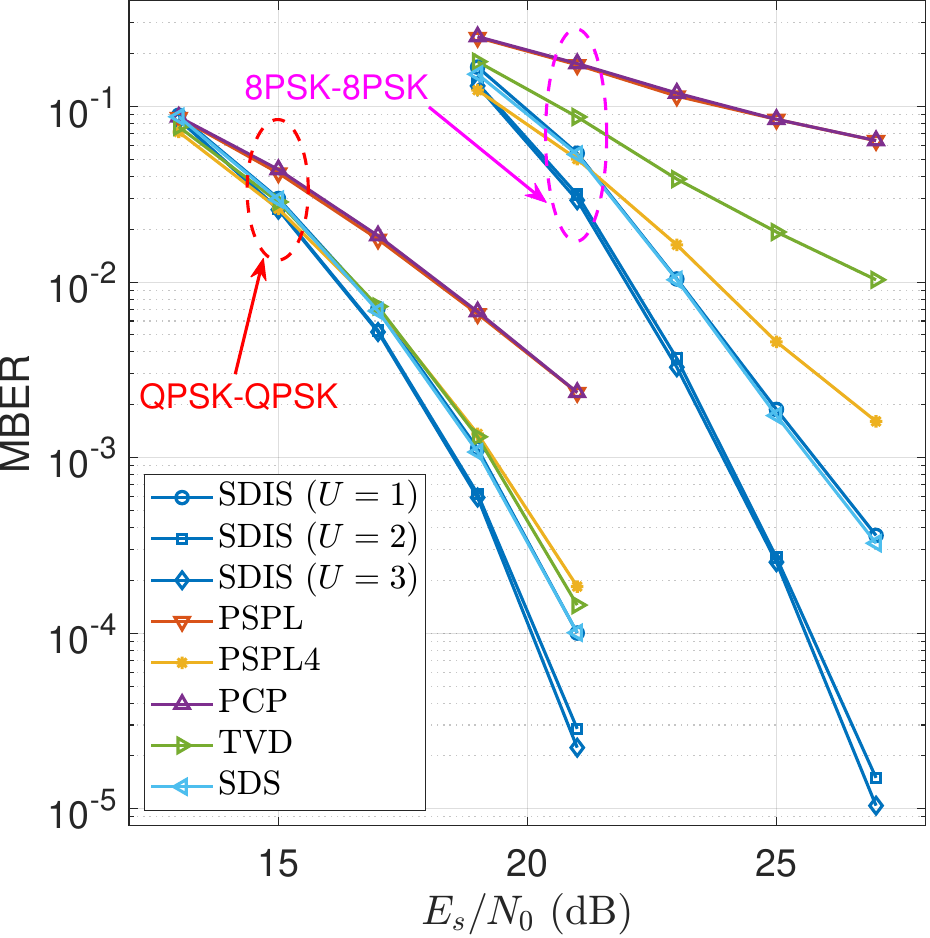}}	  	\ \ 
	\subfloat[][\label{Fig_sy_zsyfl_b}]{\includegraphics[width = 43 mm]{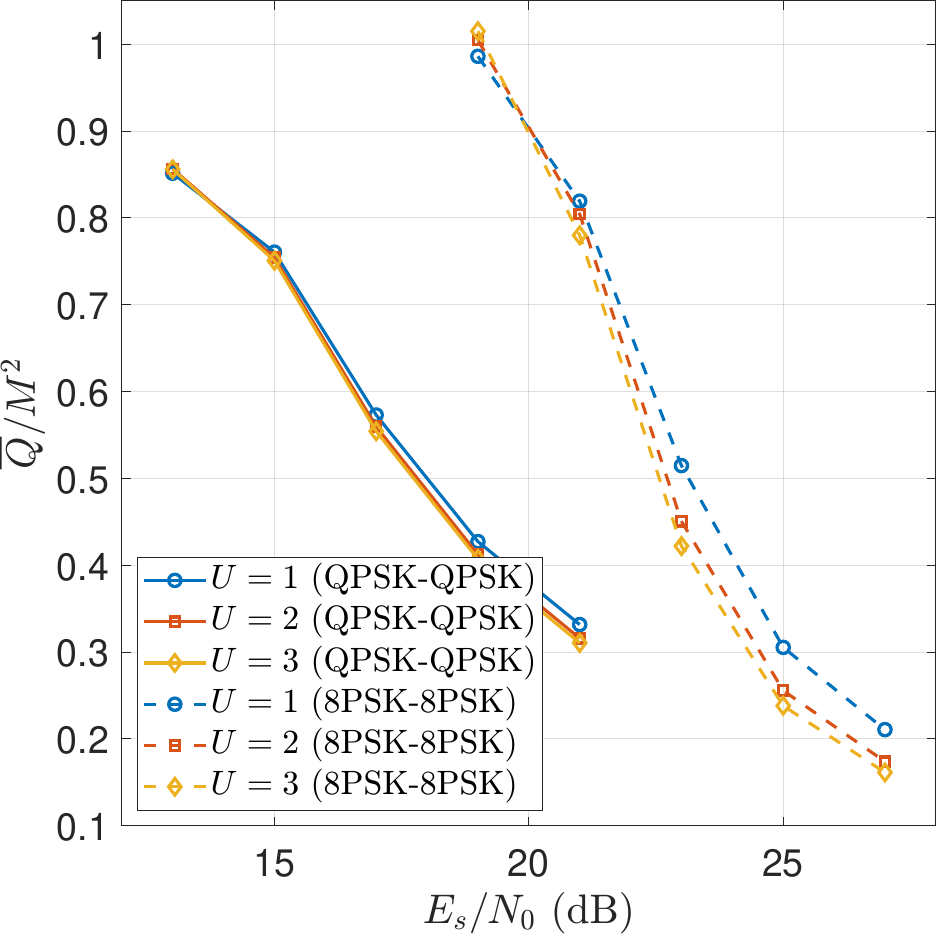}}     
	\caption{Performance of the separation algorithms under various modulation schemes and noise levels:
		(a) MBER versus $E_s / N_0$ for QPSK-QPSK and 8PSK-8PSK modulations;   
		(b) Variation of $\overline{Q}$ in SDIS with respect to noise level and modulation scheme for different $U$.}
	\label{Fig_sy_zsyfl}
\end{figure}

Fig.~\ref{Fig_sy_zsyfl} shows that SDIS achieves limited accuracy when $U = 1$, whereas decision feedback ($U > 1$) significantly improves performance.
Since the BMs are accurately estimated at $U=2$ and the signal suffers from mutual interference, the gain from increasing $U$ from $2$ to $3$ is negligible.
Correspondingly, as shown in Fig.~\ref{Fig_sy_zsyfl}(b), $\overline{Q}$ decreases as $U$ increases, owing to the PM advantage of the optimal CP that accelerates path exploration.
However, due to mutual interference, $\overline{Q}$ saturates for $U \ge 2$ and changes little thereafter.

PSPL and PCP (both $L' = 3$) achieve comparable accuracy. 
PSPL4 and TVD offer higher accuracy, but TVD's smaller $L'$ degrades its performance for 8PSK-8PSK relative to PSPL4. 
In contrast, SDIS ($U = 1$) attains an MBER comparable to that of SDS with perfectly known CSI, confirming the effectiveness of DUKF-based estimation.
SDIS ($U = 2$) employs $L' = 5$ together with DUKF for accurate channel characterization, thus yielding a significantly lower MBER than the benchmarks. 
In summary, SDIS outperforms benchmarks by at least $0.8$ dB in $E_s/N_0$ for QPSK-QPSK signals with MBER below $10^{-4}$, and by at least $2$ dB for 8PSK-8PSK signals with MBER below $3\times10^{-4}$.   

\begin{table}[t]
	\centering 
	\caption{Runtime of the Separation Algorithms (Seconds)}
	\label{Tab_sfhsyfl} 
	\begin{threeparttable}  
		\begin{tabular}{c c c c c c}
			\toprule
			Modulation / Algorithm & PSPL4 & TVD & PCP & PSPL & SDIS \\
			\midrule
			QPSK-QPSK$^\dagger$ & 1,755 	& 221 		& 108		& 106 		& 115 		\\
			8PSK-8PSK$^\dagger$ & 213,432 	& 7,139 	& 3,167 	& 3,165 	& 688 		\\ 
			\bottomrule
		\end{tabular} 
		\begin{tablenotes}[flushleft]	
			\item[$^\dagger$] $E_s/N_0 = 19$ and $25$ dB for QPSK-QPSK and 8PSK-8PSK, respectively.
		\end{tablenotes}
	\end{threeparttable}
\end{table}
 
As shown in Tables~\ref{Tab_fzd} and \ref{Tab_sfhsyfl}, the complexity of the benchmarks scales as $M^{2L'}$, making PSPL4 much slower than PSPL. 
PCP incurs slightly higher runtime due to CFO compensation, while TVD, with its two-stage processing, takes more than twice as long as PSPL. 
In contrast, SDIS scales only as $M^2$, although its DUKF is considerably more complex than the LMS used in the benchmarks. 
Consequently, SDIS runs comparably to the benchmarks for QPSK-QPSK signals, but is at least $4.60\times$ faster for 8PSK-8PSK signals.

\subsection{Impact of CSI Pre-Estimation Errors on Separation}  
\label{Sec_szjgfx_csixywcyfl}      
\begin{figure}[t]   
	\centering
	\subfloat[][\label{Fig_xywcyfl_a}]{\includegraphics[width = 43 mm]{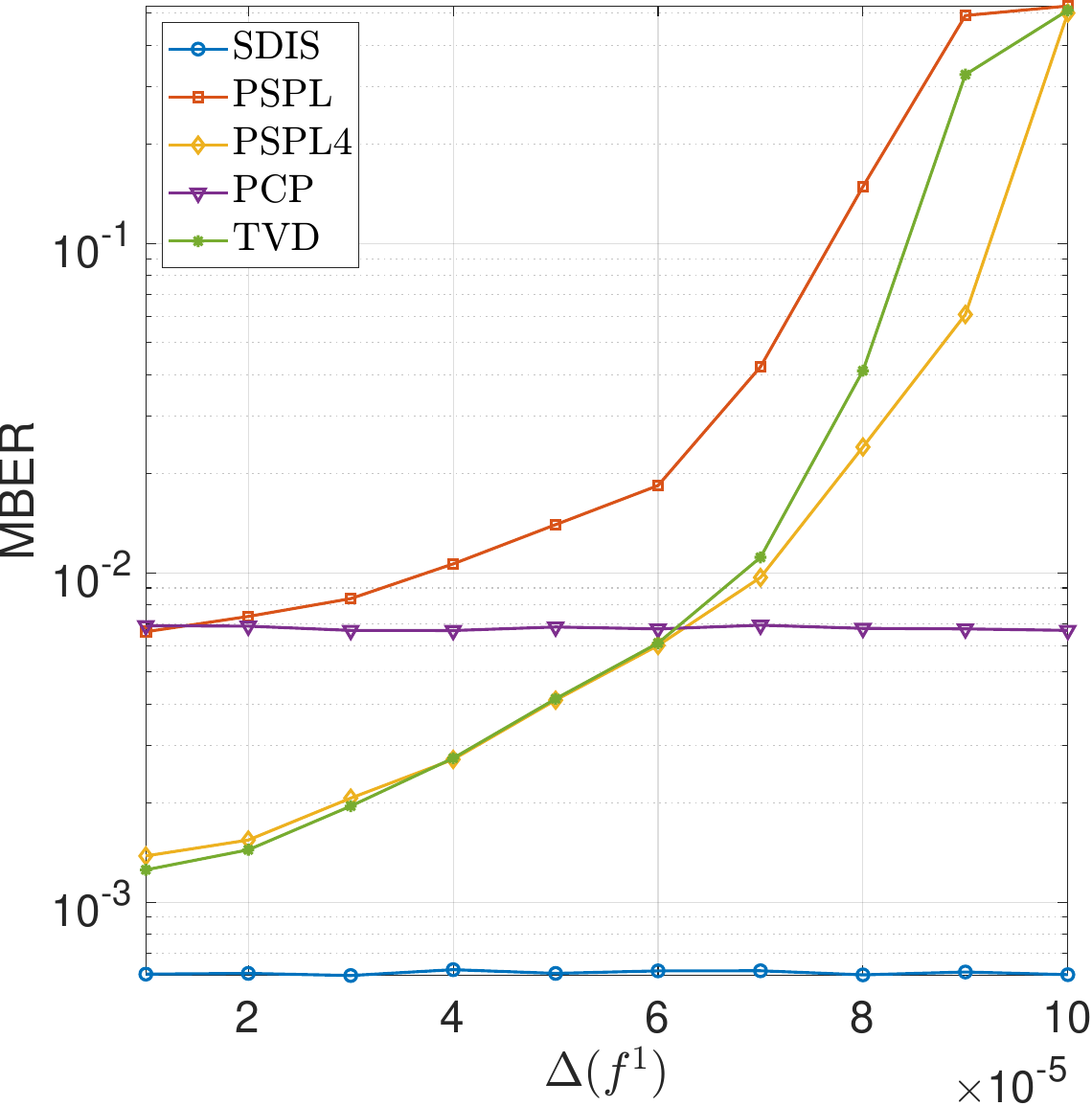}}	\ \
	\subfloat[][\label{Fig_xywcyfl_b}]{\includegraphics[width = 43 mm]{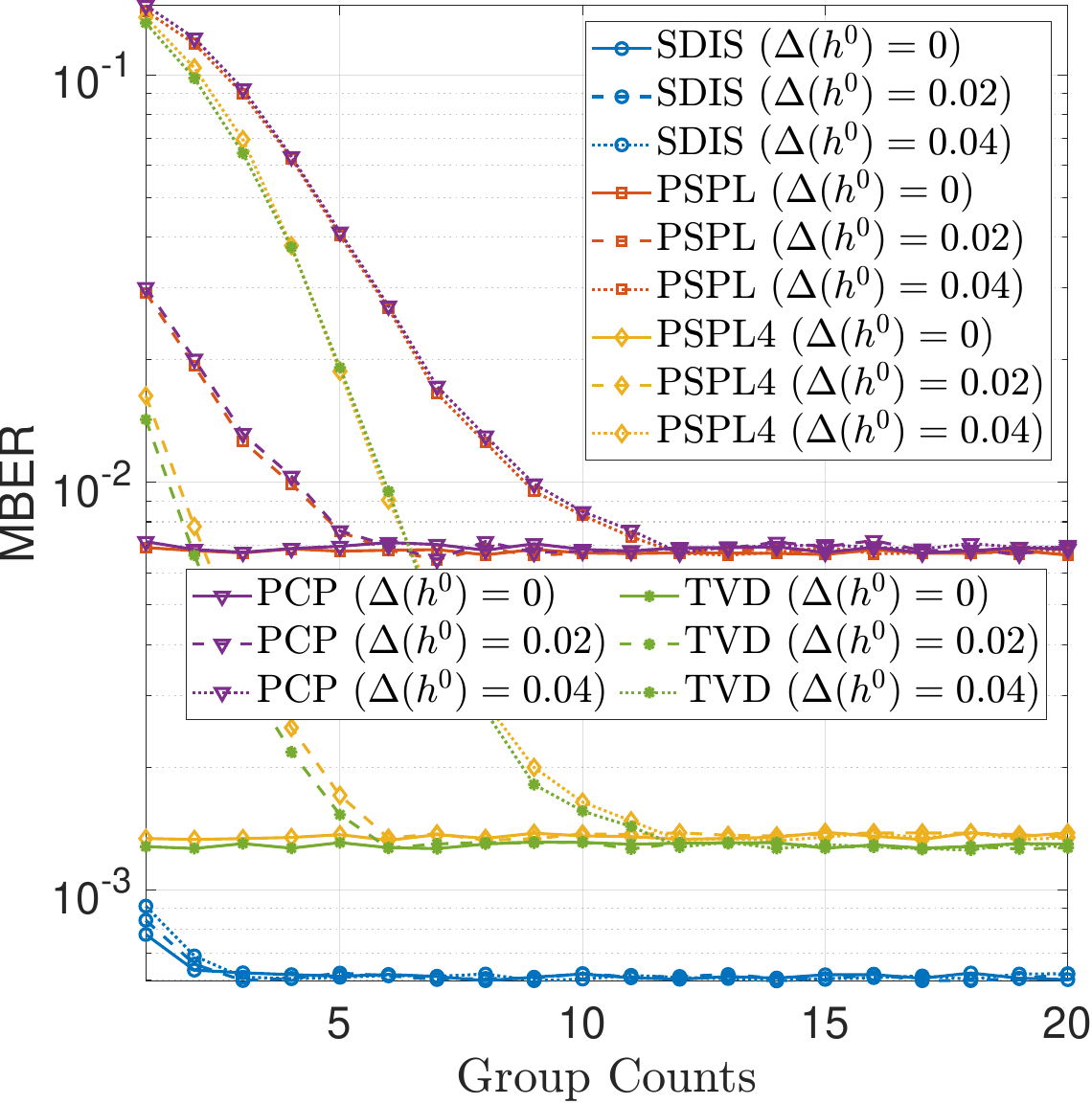}} 
	\caption{Impact of CSI pre-estimation errors on the performance of separation algorithms:
		(a) Performance under different $\Delta (f^1)$ values, where $\Delta (f^1) = -\Delta (f^0)$;
		(b) Performance under different $\Delta (h^0)$ values, where $\Delta (h^0) = \Delta (\varphi^0) = \Delta (\varepsilon^0) = -\Delta (h^1) = -\Delta (\varphi^1) = -\Delta (\varepsilon^1)$, evaluated over the first 20 data groups with 50 symbol pairs each.}
	\label{Fig_xywcyfl}   
\end{figure}  

PSPL, PSPL4, and TVD rely on LMS for CSI estimation and thus tolerate only minor CFO errors, as CFO-induced phase error accumulation degrades their performance.
As observed in Fig.~\ref{Fig_xywcyfl}(a), their MBER increases monotonically with $\Delta(f^1)$.
In contrast, PCP performs CFO pre-compensation, enabling LMS to correct the minor residual CFO-induced interference and maintain stable performance.
SDIS leverages DUKF for CSI estimation and independently tracks $\theta_n^i$ within the state vector.
Therefore, SDIS is barely affected by $\Delta(f^1)$ and consistently outperforms the benchmarks. 
  
Fig.~\ref{Fig_xywcyfl}(b) presents the MBER over the first $20$ data groups (each consisting of $50$ symbol pairs) under different $\Delta(h^i)$, $\Delta(\varphi^i)$, and $\Delta(\varepsilon^i)$.  
Due to the slow convergence of LMS, the LMS-based benchmarks require $6$ and $12$ data groups to converge at $\Delta(h^0) = 0.02$ and $0.04$, respectively.  
In contrast, SDIS employs DUKF and iterative feedback refinement, and thus achieves faster convergence and higher accuracy.  
Overall, SDIS exhibits significantly better robustness to CSI pre-estimation errors than the benchmarks.

\subsection{Impact of Noise on CSI Estimation} 
\label{Sec_szjgfx_zscsigj}

\begin{figure}[t]   
	\centering
	\includegraphics[width = 64 mm]{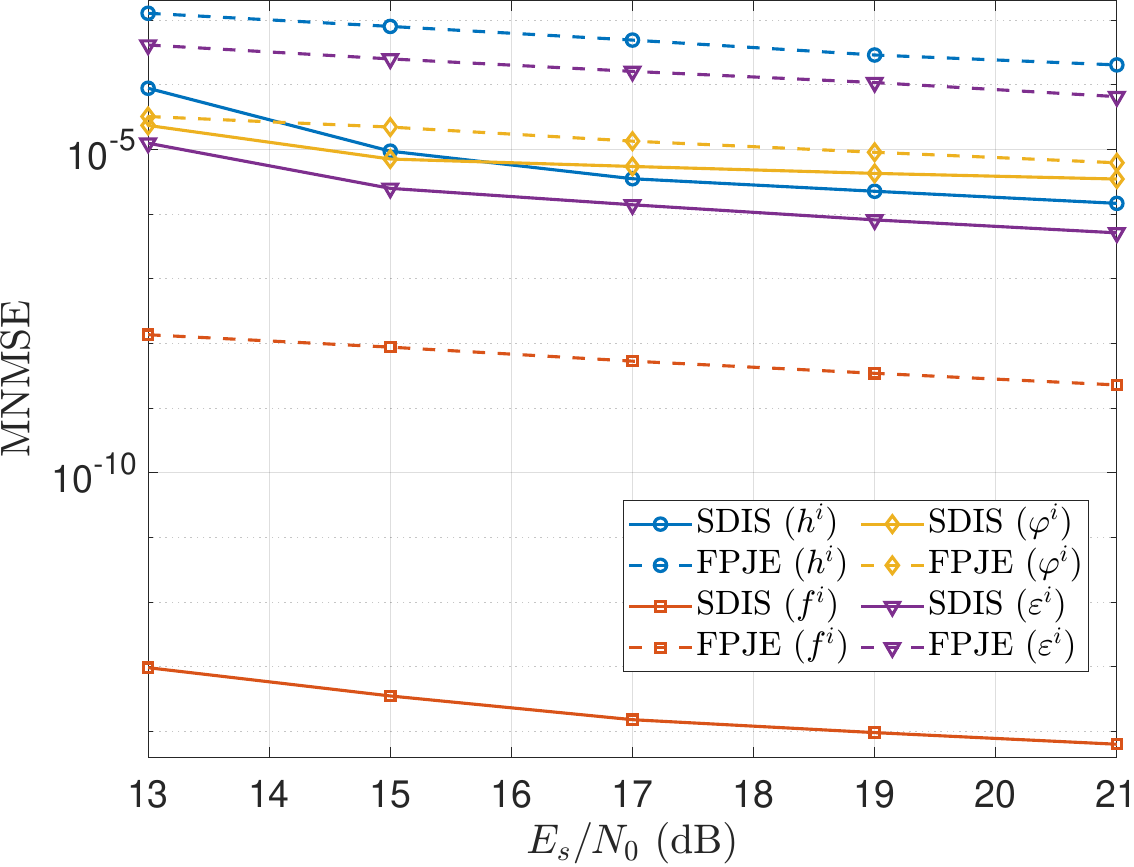}   
	\caption{Performance comparison of different CSI estimation algorithms under varying noise levels.}
	\label{Fig_zscsigj}   
\end{figure} 
  
Fig.~\ref{Fig_zscsigj} presents the performance of the CSI estimation algorithms. 
To prevent MBER variations from affecting the evaluation, we assume error-free symbol detection for FPJE. 
SDIS and FPJE both use the sequence detection results as prior information for CSI estimation. 
The difference is that FPJE performs a dimension-by-dimension stepwise search in a four-dimensional parameter space and relies on only a few dozen symbol pairs, making it prone to local optima. 
In contrast, SDIS employs DUKF for symbol-by-symbol estimation and supplements it with iterative refinement and statistical averaging to avoid local optima.
Consequently, the accuracy of SDIS for all four parameters exceeds that of FPJE. 
For CG and STO, the accuracy of SDIS is about two orders of magnitude higher than that of FPJE. 
For CFO, the MNMSE of FPJE is on the order of $10^{-9}$, whereas SDIS achieves an MNMSE on the order of $10^{-14}$.

\section{Conclusion}  
\label{Sec_jl}        
The computational complexity of conventional SCSBS algorithms scales as $M^{2L'}$. 
Owing to computational constraints, $L'$ is typically limited to $3$ in practice. 
In contrast, the proposed SDIS, which is rooted in the SD framework, detaches complexity from $L'$, thereby enabling the use of $L' = 5$ in practice.
Consequently, the combination of a larger $L'$ with iterative decision feedback significantly enhances separation accuracy. 
By incorporating DUKF for channel tracking instead of the LMS used in the benchmarks, SDIS achieves superior CSI estimation accuracy and exhibits robust performance against CSI pre-estimation errors.
Overall, SDIS exhibits significant advantages in separation accuracy, CSI estimation precision, computational efficiency, and robustness. 
To achieve an MBER below $3 \times 10^{-4}$ for 8PSK-8PSK signals, SDIS is at least $4.60$ times faster than the benchmarks, while also requiring at least $2$ dB less $E_s/N_0$. 
Future work will extend this research to scenarios involving more than two sub-signals.

\bibliographystyle{IEEEtran}                                      
\bibliography{IEEEabrv, Corr_Ref_31}

\begin{thebibliography}{10}
\providecommand{\url}[1]{#1}
\csname url@samestyle\endcsname
\providecommand{\newblock}{\relax}
\providecommand{\bibinfo}[2]{#2}
\providecommand{\BIBentrySTDinterwordspacing}{\spaceskip=0pt\relax}
\providecommand{\BIBentryALTinterwordstretchfactor}{4}
\providecommand{\BIBentryALTinterwordspacing}{\spaceskip=\fontdimen2\font plus
\BIBentryALTinterwordstretchfactor\fontdimen3\font minus
  \fontdimen4\font\relax}
\providecommand{\BIBforeignlanguage}[2]{{%
\expandafter\ifx\csname l@#1\endcsname\relax
\typeout{** WARNING: IEEEtran.bst: No hyphenation pattern has been}%
\typeout{** loaded for the language `#1'. Using the pattern for}%
\typeout{** the default language instead.}%
\else
\language=\csname l@#1\endcsname
\fi
#2}}
\providecommand{\BIBdecl}{\relax}
\BIBdecl

\bibitem{r_nhw_2025}
H.~Niu, L.~Wang, Z.~Lu, C.~Wu, and X.~Wen, ``{SINR}-adaptive and
  {CBR}-controllable semantic cellular communication considering imperfect
  {CSI} and inter-cell co-channel interference,'' \emph{{IEEE Trans. Cogn.
  Commun. Netw.}}, vol.~11, no.~4, pp. 2368--2385, Aug. 2025.

\bibitem{r_djz_2025}
J.~Ding, L.~Wei, Y.~Zhao, and B.~Jiao, ``Exceeding spectral efficiency gain of
  2 with co-frequency co-time full-duplex in finite blocklength regime,''
  \emph{{IEEE Commun. Lett.}}, vol.~29, no.~7, pp. 1619--1623, Jul. 2025.

\bibitem{r_wh_2025}
H.~Wang, K.~Gong, L.~Zhang, W.~Wang, X.~Wei, P.~Sun, and H.~Jiang, ``Blind
  parameters estimation for time-frequency overlapped satellite signals,''
  \emph{{IEEE Commun. Lett.}}, vol.~29, no.~6, pp. 1486--1490, Jun. 2025.

\bibitem{r_tsl_2008}
S.~Tu, H.~Zheng, and N.~Gu, ``Single-channel blind separation of two {QPSK}
  signals using per-survivor processing,'' in \emph{{Proc. APCCAS 2008}}, Dec.
  2008, pp. 473--476.

\bibitem{r_lxb_2017}
X.~Liu, Y.~Guan, S.~N. Koh, Z.~Liu, and P.~Wang, ``Single-channel blind
  separation of co-frequency {PSK} signals with unknown carrier frequency
  offsets,'' in \emph{{Proc. MILCOM 2017}}, 2017, pp. 641--646.

\bibitem{r_yy_2020}
Y.~Yang, H.~Peng, D.~Zhang, and P.~Wang, ``Iterative processing structure for
  the single-channel mixture of digital-modulated adjacent-frequency source
  signals,'' \emph{{IEEE Trans. Veh. Technol.}}, vol.~69, no.~2, pp.
  1639--1650, Feb. 2020.

\bibitem{r_gwb_2022}
W.~Guo, H.~Zhao, C.~Song, S.~Shao, and Y.~Tang, ``Direct-link interference
  cancellation design for backscatter communications over ambient {DVB}
  signals,'' \emph{{IEEE Trans. Broadcast.}}, vol.~68, no.~2, pp. 317--330,
  Jun. 2022.

\bibitem{r_gym_2019}
Y.~Guo and H.~Peng, ``Single channel blind separation performance bound of
  non-cooperative received paired carrier multiple access mixed signal,''
  \emph{{J. Electron. Inf. Technol.}}, vol.~41, no.~01, pp. 240--247, Jan.
  2019.

\bibitem{r_my_2}
\BIBentryALTinterwordspacing
H.~Wang, K.~Gong, P.~Sun, W.~Wang, and H.~Jiang, ``Low-complexity sequential
  detection framework for single-channel co-frequency signal separation,''
  2026. [Online]. Available: \url{https://arxiv.org/abs/2609.03986}
\BIBentrySTDinterwordspacing

\bibitem{r_fa_2020}
A.~Feder, W.~Wicke, M.~Hirschbeck, and W.~Gerstacker, ``Blind symbol rate and
  frequency offset estimation for {PCMA} signals via cyclic correlations,'' in
  \emph{{Proc. GLOBECOM 2020}}, 2020, pp. 1--7.

\bibitem{r_fa_2021}
A.~Feder, W.~Gerstacker, and M.~Hirschbeck, ``Blind symbol timing and carrier
  phase estimation for {PCMA} satellite signals via cyclic statistics,'' in
  \emph{{Proc. GLOBECOM 2021}}, 2021, pp. 1--7.

\bibitem{r_lmq_2024}
M.~Liu, S.~Yu, Y.~Chen, and S.~Chen, ``Blind parameter estimation for
  co-channel digital communication signals,'' \emph{{Wirel. Netw.}}, vol.~30,
  no.~6, pp. 5589--5599, Aug. 2024.

\bibitem{r_gym_2018}
Y.~Guo, H.~Peng, and J.~Fu, ``Joint blind parameter estimation of
  non-cooperative high-order modulated {PCMA} signals,'' \emph{{KSII Trans.
  Internet Inf. Syst.}}, vol.~12, no.~10, pp. 4873--4888, Oct. 2018.

\bibitem{r_by_2025}
Y.~Berrouche, M.~Kulhandjian, and H.~Kulhandjian, ``A hyperbolic secant-based
  pulse for enhanced {FTN} signaling in {5G/6G} systems,'' \emph{{IEEE Wirel.
  Commun. Lett.}}, vol.~14, pp. 380--384, Oct. 2025.

\bibitem{r_lhy_2025}
H.~Liu, X.~Sun, J.~Yang, M.~Xu, and S.~Bai, ``Skewed unscented {Kalman} filter
  using {Gaussian} sum,'' \emph{{IEEE Trans. Aerosp. Electron. Syst.}},
  vol.~61, no.~2, pp. 3917--3935, Apr. 2025.

\end{thebibliography}

\end{document}